\def\year{2017}\relax
\documentclass[letterpaper]{article} 
\usepackage{aaai17}  
\usepackage{times}  
\usepackage{helvet}  
\usepackage{courier}  
\usepackage{url}  
\usepackage{graphicx}  
\expandafter\def\expandafter\UrlBreaks\expandafter{\UrlBreaks
  \do\a\do\b\do\c\do\d\do\e\do\f\do\g\do\h\do\i\do\j%
  \do\k\do\l\do\m\do\n\do\o\do\p\do\q\do\r\do\s\do\t%
  \do\u\do\v\do\w\do\x\do\y\do\z\do\A\do\B\do\C\do\D%
  \do\E\do\F\do\G\do\H\do\I\do\J\do\K\do\L\do\M\do\N%
  \do\O\do\P\do\Q\do\R\do\S\do\T\do\U\do\V\do\W\do\X%
  \do\Y\do\Z}
\usepackage{floatrow}
\usepackage[inline]{enumitem}
\usepackage{multirow}
\DeclareFloatFont{small}{\small}
\begin{document}
%
 \title{The Impact of Crowds on News Engagement: A Reddit Case Study}

\author{ 
Benjamin D. Horne and Sibel Adal{\i} \\
Rensselaer Polytechnic Institute\\
110 8th Street, Troy, New York, USA\\
\{horneb, adalis\}@rpi.edu
}

\maketitle
\begin{abstract}
Today, users are reading the news through social platforms. These platforms are built to facilitate crowd engagement, but not necessarily disseminate useful news to inform the masses. Hence, the news that is highly engaged with may not be the news that best informs. While predicting news popularity has been well studied, it has not been studied in the context of crowd manipulations. In this paper, we provide some preliminary results to a longer term project on crowd and platform manipulations of news and news popularity. In particular, we choose to study known features for predicting news popularity and how those features may change on \url{reddit.com}, a social platform used commonly for news aggregation. Along with this, we explore ways in which users can alter the perception of news through changing the title of an article. We find that news on reddit is predictable using previously studied sentiment and content features and that posts with titles changed by reddit users tend to be more popular than posts with the original article title.
\end{abstract}

\section{Introduction}
It is well-known that an operational and useful democracy relies on an educated and well-informed population~\cite{lewandowsky2012misinformation}. The primary way this population is informed is through the news. While traditionally, news is received through print, television, or online articles, increasingly more people consume news through social platforms. On these platforms, both crowds and the system work together to makes decisions on what information is important. These decisions may be informed by many factors such as homophily, estimated interests, self-selection of sharing, or the crowd voting on the information. Recently, many researchers have begin studying the system side impact of social platforms on information engagement~\cite{bakshy2015exposure}~\cite{horne2016impact}, but very few studies have looked at the crowds' effect on these information decisions, especially in the context of news. If we understand both the impact of the platform and the impact of the platform user on news engagement and popularity, we can better build social platforms that not only engage users, but better inform users.

One such social platform is \url{reddit.com}. In the context of news, reddit is a platform where users post links to news articles and vote on their importance or relevance. The number of votes will roughly determine the order in which news is sorted on the page. When users post links, they have the option to use the title of the article they are linking to, or to create their own. In addition, users can comment on the news article posts to start a discussion with the community. Hence, on reddit, crowds can influence news consumption through crowd selection, voting, title changing, and direct discussion.

In this paper, we will report preliminary findings on three of these crowd effects: voting, commenting, and title changing. In particular, we will ask two questions: \begin{enumerate*}[label=(\arabic*)]\item Can known features from the news popularity literature predict the number of votes and the number of comments on reddit news posts?\item Does the crowd affect the number of votes and number of comments through changing the news headlines? \end{enumerate*} To answer these questions, we will compute features and predict popularity on scraped news articles posted to the reddit community: \texttt{r/worldnews}. After assessing our prediction models, we will introspect on our features to gain further insight into how each important feature predicts news popularity on reddit. Lastly, we will perform pairwise hypothesis testing on news media made titles and reddit user made titles to gain insight into one specific way the crowd can manipulate news.  

We find that both the number of votes and the number of comments on reddit news posts is predictable using previously known, non-temporal news popularity features. We find that votes (also known as the score) are predicted well by a mixture of content, sentiment, and subjectivity features, whereas the number of comments is predicted well by using only sentiment and subjectivity features. Hence, the motivation to vote on an article and the motivation to engaged with an article are different. Specifically, voting is based on content quality, while commenting is based solely on emotion. In addition, we find that post titles that are changed from the original news article title tend to receive slightly higher scores and slightly more comments than posts using the original news article title. When titles are changed, they are changed to be significantly more positive, less negative, more informal, more difficult to read, and longer overall. These findings illustrate the reddit community's ability to make news headlines that engage their audience by adding their own positive analysis to the headline. 

\section{Related Work}
\subsection{News Popularity}
News and information popularity is well-studied. We know that, in general,  shared news tends to be more negative, and that people tend to share information that will get an emotional response~\cite{harcup2001news}~\cite{lewandowsky2012misinformation}. One recent study demonstrated that headline negativity and overall sentiment is important in news popularity.~\cite{reis2015breaking}. The authors extract sentiment features from the headlines of four major news sources and use the bit.ly API to infer popularity. The study shows that the majority of news produced negative headlines and this negativity is fairly constant over time. Along with this, extreme sentiment on both the positive and negative side tends to attract more popularity. To test these finding in our study, we will use the same sentiment tool, SentiStrength~\cite{thelwall2010sentiment}, to evaluate the intensity of sentiment in highly engaged news. In a similar study, Keneshloo et al.'s work, on predicting the popularity of news articles in the Washington Post~\cite{keneshloo2016predicting}, shows that only the neutrality of content is an important sentiment predictor of popularity. However, the authors do not give us any insight as to if high neutrality means high popularity or if low neutrality means high popularity. To test this finding in our study, we will also use the sentiment tool used in Keneshloo et al.'s work, Vader-Sentiment~\cite{hutto2014vader}, to compute the positive, negative, neutral, and composite (overall) sentiment. Keneshloo et al. also show that readability using the SMOG index and the title length are important content features in popularity prediction. Hence, we will also implement these features in our study. 

From a traditional journalism perspective, there exists what are called news values.~\cite{harcup2001news} While these values differ slightly between authors, we will implement features on some commonly held news values, including simplicity, sentiment, surprise, and unambiguous arguments.

Also, related to this study, is the large body of work on general information popularity prediction. These studies include predicting retweets on Twitter~\cite{zaman2014bayesian} and the number of comments on a news article~\cite{tsagkias2009predicting}. Some of our features will overlap with these general popularity studies.

\subsection{News Headlines}
There has also been significant work on the influence of news headlines. While on the surface changing a title seems trivial, it can have an substantial impact on how people perceive and consume news. In a 2014 study, Ecker et al. found that misleading news titles can emphasize secondary content rather than the primary content of the article, and that misleading titles affect the readers memory, reasoning, and intentions, as they struggle to update their memory to more exact information~\cite{ecker2014effects}. Furthermore, in a 2007 study by Surber and Schroeder, titles were found to improve recall of important information~\cite{surber2007effect}. If title information is false or even slightly misleading, this information recall will be flawed. To make matters worse, it is well known that humans take shortcuts in information trust decisions when there is a low need for cognition, low energy, or high cognitive strain~\cite{petty1986}. Hence, users are prone to formulate opinions about the news simply from the title. This effect may be multiplied on platforms such as reddit, as users may be prone to voting on information without fully exploring the news article. Therefore, we will explore the impact of many different features on the both the titles of the articles and the titles of the reddit posts.

\subsection{Reddit}
Reddit, as a platform, has also been very well studied. In particular, we know that the posts' timing and the posts' titles are important factors in popularity on reddit~\cite{lakkaraju2013s}~\cite{tran2016characterizing}, but this popularity can be different across communities~\cite{jaech2015talking}. In addition, we know that reddit user reputation has little influence on popularity, unlike on Twitter where users are not anonymous.

There has also been numerous studies that explore the
higher-level behavior of reddit, and the many confounding factors in popularity on reddit. Hessel, Tan, and Lee study the nature of ”spin-off” or highly-related communities
on reddit using a large comprehensive data set that
spans over 8 years~\cite{hessel2016science}. They show that users who explored new ``spin-off" communities become more active in the original community rather than the spin-off community they explored. In Hessel, Lee, and Mimno's 2017 work, they find that visual and text features together predict popularity better than author-based features across several image-based communities~\cite{hessel2017cats}. They again show that popularity is influenced by timing factors including the time of day.  Gilbert shows that reddit overlooks about 52\% of the most popular links at first submission~\cite{gilbert2013widespread}. This early finding is likely due to the recently found popularity factors such as the timing of posts and the noisy sorted order on the page, which is only roughly reflected by the community votes. Further, these factors can cause ``rich-get-richer" senerios, where already popular posts continue to gain popularity, while not so popular posts continue to get overlooked~\cite{hessel2017cats}. All of these studies provide insight into the confounding factors in reddit popularity that may impact our predictions: timing, user activity, sorting displayed, noise in scores, community differences, and author popularity.

Since timing is both a confounding factor in reddit popularity and in general news popularity, one might expect that non-temporal features will have very little predictive power. We will investigate this further in this paper to check if the content of news still matters in determining its popularity above and beyond its timing. Performing this same analysis with time controlled data is left for our future work.

\section{Methods}
\subsection{Data Sets}
We extract posts from 2012 and 2013 for one popular news community on reddit: \texttt{r/worldnews}. Once we extract all posts, we extract the voting score, number of comments, post title, and news story urls from each post. These news story urls are used to scrape a sample of news articles, including the body text and title text, using a mix of our own code and the Python Goose library. We will filter out any article that is under 100 characters or blocked by a paywall. This reddit data comes from Tan and Lee's reddit post data set~\cite{tan2015all} and Hessel et al.'s full comment tree extension to that reddit dataset~\cite{hessel2016science}. Both of these data sets are based on a reddit API collection originally done by Jason Baumgartner of \url{pushshift.io}. Our newly collected data set of news articles with corresponding reddit engagement statistics can be found at \url{https://github.com/rpitrust/reddit-scraped-worldnews-dataset}.

\begin{figure}
\begin{center}
  \begin{tabular}{cc} \\
    {\small Score} & {\small Comments} \\
    \includegraphics[width=100pt,keepaspectratio=true]{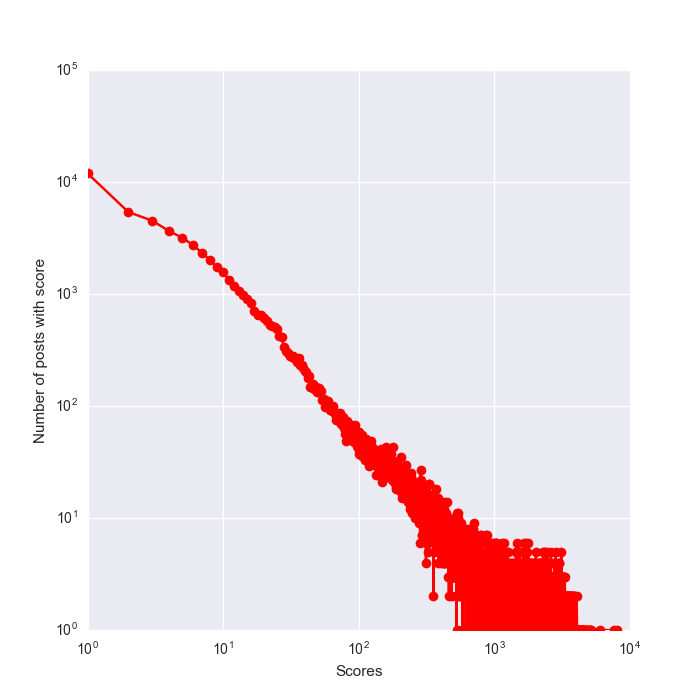}
    & \includegraphics[width=100pt,keepaspectratio=true]{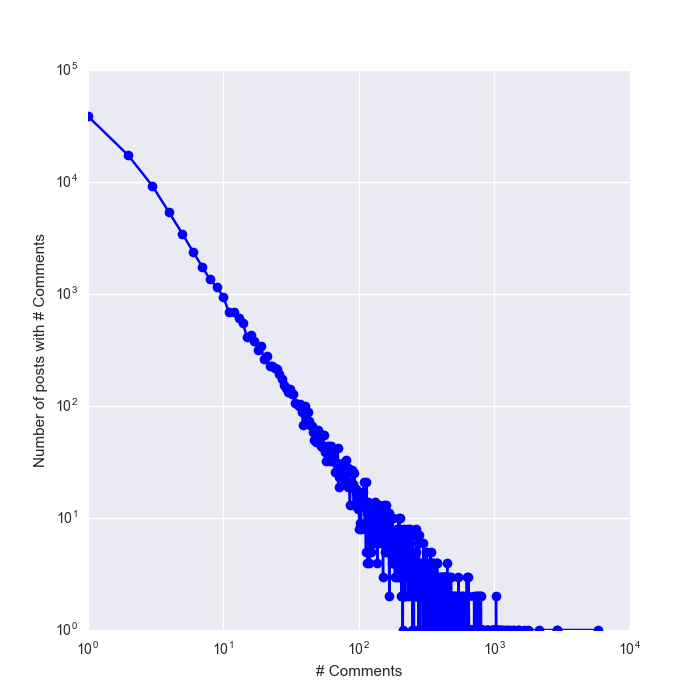} \\
    \multicolumn{2}{c}{(a) All Data} \\
    {\small Score} & {\small Comments} \\
      \includegraphics[width=100pt,keepaspectratio=true]{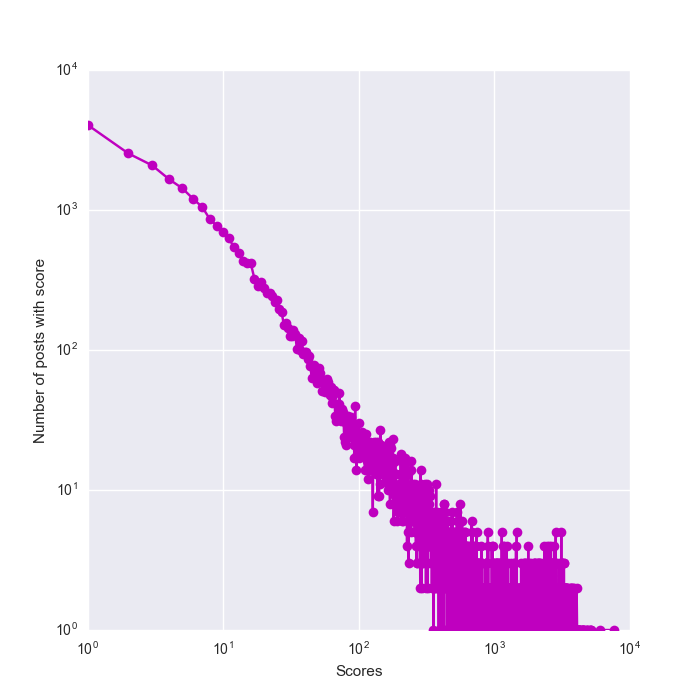}
      & \includegraphics[width=100pt,keepaspectratio=true]{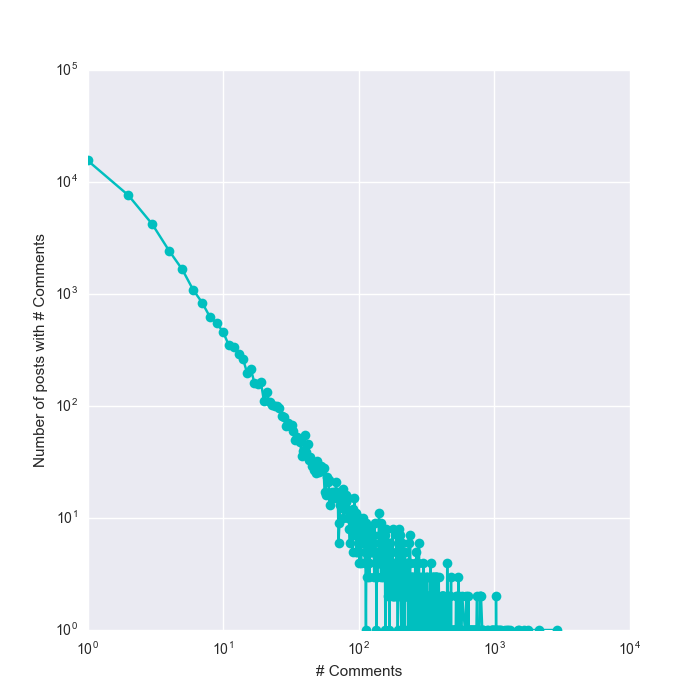} \\
      \multicolumn{2}{c}{(b) Scraped Data} \\
\end{tabular}
  \caption{The \textbf{Score} and \textbf{\# of Comment} distributions on teh full reddit data set and on the article scraped data set for \texttt{r/worldnews} 2013. The distributions are similar for \texttt{r/worldnews} 2012. \label{dists}}
\end{center}
\end{figure}

The number of posts extracted and number of articles scraped in each year can be seen in Table~\ref{dataids}. In Figure~\ref{dists}, we show the distribution of scores and distribution of number of comments for the full reddit data set and for the articles scraped. As expected, both distributions are power-law (or more precisely Zipfian) distributions~\cite{jaech2015talking}, and our scraped data set provides an adequate random sample of the distributions for both the score and the number of comments. 

\begin{table}[h]
\centering
\begin{tabular}{ | c | c || c | c | }
\hline
 subreddit & year & \# posts & \# news articles scraped \\
 \hline
 r/worldnews & 2012 &  60734 &  31938\\
 r/worldnews & 2013 &  93254 &  40809\\
 \hline
 \textbf{total} & & \textbf{154K} & \textbf{72.7K}\\
 \hline
 \end{tabular}
 \caption{\label{dataids} The number of posts and scraped articles per data set in this study.}
 \end{table}

\begin{table*}[thb!]
\centering
\begin{minipage}[t]{3.2in}
\begin{tabular}{p{0.6in}p{2.2in}} 
Abbr. & Description \\ \hline 
  str\_neg & negative emotion using SentiStrength\\
  	str\_pos & positive emotion using SentiStrength\\
  	vad\_neg & negative sentiment score Vader-Sentiment\\
  	vad\_pos & positive sentiment score Vader-Sentiment\\
  	vad\_neu & neutral sentiment score Vader-Sentiment\\
  	vad\_comp & composite sentiment score Vader-Sentiment\\
  	NB\_psubj & probability of subjectivity using a learned Naive Bayes classifier\\
  	NB\_pobj & probability of objectivity using a learned Naive Bayes classifier\\
  	NB\_subjcat & binary category of objective or subjective\\
  	posemo & number of positive emotion words LIWC\\
  	negemo & number of negative emotion words LIWC\\
  	tone & number of emotional tone words\\
  	affect & number of emotion words (anger, sad, etc.)\\
	analytic & number of analytic words\\  	
  	insight & number of insightful words\\
  	authent & number of authentic words \\
  	tentat & number of tentative words \\
  	certain & number of certainty words \\
  	affil & number of affiliation words \\
  	focuspresent & number of present tense words\\
  	focusfuture & number of future tense words\\
  	&\\
  	&\\
  	\hline \\
\multicolumn{2}{c}{(a) Sentiment and Subjectivity Features}
\end{tabular}
  \end{minipage} 
\begin{minipage}[t]{3.2in}
\begin{tabular}{p{0.6in}p{2.4in}}  
Abbr. & Description \\ \hline
  WC & word count\\
  WPS  & words per sentence\\
  GI & Gunning Fog Grade Readability Index\\
  SMOG & SMOG Readability Index\\
  FK & Flesh-Kincaid Grade Readability Index\\
  flu\_coca\_c & avg. frequency of least common 3 words using all of the coca corpus \\
  flu\_coca\_d & avg. frequency of words in each document using all of the coca corpus \\
   TTR & Type-Token Ratio (lexical diversity) \\
   avg\_wlen & avg. length of each word \\
   quote & number of quotation marks\\
   ppron & number of personal pronouns\\
    i & number of I pronouns\\
  we & number of we pronouns \\
  you & number of you pronouns\\
  shehe & number of she or he pronouns\\
  quant & number of quantifying words\\
  swear & number of swear words \\
  netspeak & number of online slang terms (lol, brb)\\
  interrog & number of interrogatives (how, what, why)\\
  per\_stop & percent of stop words (the, is, on)\\
  allPunc & number of punctuation\\
  quotes & number of quotes\\
  function & number of function words \\
  word\_len & average word length \\
  \hline \\
  \multicolumn{2}{c}{(b) Content Features} \\ \\
\end{tabular}
\end{minipage}
\caption{\label{tbl:features} Different features used in our study}
\end{table*}

\subsection{Features}
We engineer features with two questions in mind: \begin{enumerate*}[label=(\arabic*)] \item What do we already know about popular news? \item In what ways may reddit crowds change what news is popular?  \end{enumerate*} From this perspective, we implement many features from the popularity literature and some features of our own. These features can be divided into three main classes: sentiment, subjectivity, and content structure. We will provide detailed descriptions of the more complex features here. The complete feature set can be found in Table~\ref{tbl:features}. 

To compute sentiment features, we will use 3 different approaches from literature: SentiStrength, Vader-Sentiment, and LIWC. SentiStrength is a machine learning based method that provides the polarity of sentiment, with -5 being very negative and +5 being very positive~\cite{thelwall2010sentiment}. In previous work, it has been shown to work well with news data~\cite{reis2015breaking}. Vader-Sentiment is a lexicon and rule-based sentiment tool that is built for sentiment expressed on social media~\cite{hutto2014vader}. It provides measures of positive, negative, and neutral sentiment, along with a composite measure to provide a single overall measure of sentiment. Vader-Sentiment was found to work well on news data in previous work~\cite{keneshloo2016predicting}. Lastly, Linguistic Inquiry and Word Count (LIWC) is a well known bag of words (or dictionary) based method for measuring many different psychological and language dimensions~\cite{tausczik2010psychological}. We will specifically use the positive emotion, negative emotion, tone, and affect word counts to measure sentiment and emotion. Along with sentiment, we will compute several features that capture the general subjectivity of a news article or title. Primarily, we will use a Naive Bayes classifier which we have trained on a data set of 10K labeled subjective and objective sentences from Pang and Lee's 2004 work~\cite{pang2004sentimental}. The classifier achieves 92\% 5-fold cross-validation accuracy. We will use the classifier to create three features: the probability the text is objective, the probability the text is subjective, and a hard binary classification of objective or subjective. While this data set has been used in many different scenarios, to our knowledge it has not been used on news data. In addition to these features, we provide several LIWC features that capture something about the objectivity of the articles, including analytic, insight, authentic, tentative, certain, affiliation, focus-present, and focus-future word counts.

Additionally, we will provide simple content features to capture some other findings in the news popularity literature and general information trust literature. These include readability using three different well-known readability metrics (SMOG, Gunning-Fog, Flesh-Kincaid), a measure of lexical diversity (Type-Token Ratio), and a set of LIWC language dimension features. In addition, we will compute a feature called fluency from Horne et al.~\cite{horne2016expertise}. Fluency is a measure of word commonality and technicality. It is computed by counting the frequency of
each word appearing in a large English corpus. If a piece of
text has a high fluency score, we would expect the words being
used are more common words. On the other hand, if the
fluency score is low, we would expect many of the words to
be highly technical or rare. We choose compute
the feature using the Corpus of Contemporary American
English (COCA)~\cite{coca}. While fluency has not been explicitly tested in news popularity scenarios, it is meant to capture the well known information processing concept of perceived familiarity and coherence of information. It is known that people may determine the believability of information by how coherent that information is cognitively. Information that is familiar or has been seen before tends to be more coherent, even if that information is false~\cite{lewandowsky2012misinformation}. 

\subsection{Testing News Popularity Findings}
To test our feature sets and what we already know about popular news, we will use the following methodology:

\begin{enumerate}
\item Learn models to predict the score of a news article in terms of various combinations of feature sets.
\item Learn models to predict the number of comments under a news article in terms of various combinations of feature sets.
\item Assess all models on the task of ranking.
\end{enumerate}

Since, each reddit community is set up as a ranking of posts based on voting (and some unknown noise to keep the system from being manipulated), it is natural for us to evaluate our models on the task of ranking.

\subsubsection{Learning to Rank}
To perform learning to rank, we will use the Lasso regression algorithm from the Python scikit-learn library~\cite{pedregosa2011scikit}, with the fitting and regularization parameters being chosen by cross-validation. Lasso is a linear model that performs L1 regularization and provides sparse solutions. It is known for its ability to handle a large number of correlated features well.

\subsubsection{Preprocessing}
We will perform one preprocessing step on each data set, in which all scores or number of comments that fall below 30th percentile will be removed. Since the distributions are fat-tailed, these removed posts roughly have scores of 1 or below and number of comments of 0. We perform this step to ensure we are learning the difference between the engagement of actual news and not spam submitted to the community. While our news scraping methodology removes much of the spam, we want to ensure our learning algorithm is not skewed towards spam, creating a different problem of separating spam from news. Choosing the 30th percentile is simply a heuristic to remove posts with scores and number of comments of 1 or 0.

\subsubsection{Learning to Rank Metrics}
To evaluate the performance of our models, we will first divide the data into a train and test set with 80\% of the data being for training and 20\% of the data being for testing. Once the models are learned, we evaluate the performance on the test set by the following metrics from the learning to rank literature:
\begin{enumerate}
\item Precision @ k: Percentage of top k posts we were able to
retrieve correctly.
\item Kendall-tau distance (KT-distance) @ k: Kendall-tau distance between the relative ranking of the
top k posts according to the real score versus the relative
ranking of the same k posts by their predicted scores.
\end{enumerate}

We will report $k = 3, 10, 100, 500$. We use both precision and Kendall-tau distance to gain a complete picture of our prediction quality. If we achieve high precision on large test sets and the Kendall-tau distance is low, it means that the top k posts were correctly retrieved and the relative
ranking of the top posts are close to the real ranking.
This allows us to conclude that the predicted ranking is close
to the real ranking overall. A rule of thumb for assessing these results is to have a high precision at low values of $k$ and a low Kendall-tau distance at high values of $k$.

\subsubsection{Feature Introspection}
To better understand how our features predict news determined popular by the crowd, we will perform a two step post-hoc analysis methodology. First, we will perform stability selection (implemented in the Randomized Lasso class of Python Scikit-learn) to select the most important features used in the prediction of each model. Second, we will perform traditional hypothesis testing using Wilcox Rank-Sum tests on a two-class divide of those selected important features. This step will describe the shift in each independent feature distribution, thus, helping describe exactly what ways highly popular news and not highly popular news differ. The data will be divided into a high class (any story with a score or number of comments above the 90th percentile) and a low class (any story with a score or number of comments below the 50th percentile). These divides are simply heuristics for capturing the top and bottom of our heavily skewed distributions. Once again, our goal in creating this divide is to provide a better understanding of how our features are predicting, not just which features are important. We will report the direction each features shifts between the two-classes and if the shift is statistically significant.

\subsection{Exploring Title Change}
Additionally, we find that users on reddit tend to significantly edit the titles that they post. In 85\% of the posts in 2012 and 72\% of posts in 2013, titles have been edited by at least one word. We would like to study both how the titles are changed and what effect these changes have on popularity of the posts. To do this, we will first compute the cosine similarities between each reddit user title and the original title from the news source. This will determine the distribution of title changes. Second, we will determine if changing the title has any influence on the score and number of comments a news article receives by using traditional hypothesis testing on a 2 class divide as described previously. We will also report several signficance test measures on this divide, including the Wilcox Rank-Sum p-values, the Cohen's d effect size, and the Grissom-Kim probability. The Grissom-Kim probability is simply the probability that a randomly selected number in group A is greater than a randomly selected number in group B. Grissom and Kim propose this method to check the effect size on nonparametric tests, as Cohen's d may not always be accurate on heavily skewed distributions~\cite{grissom2012effect}. Reporting both metrics will provide a complete picture of the titles in this data set. Finally, we will use Wilcox Rank-Sum tests on extracted title pairs that have a similarity less than $0.1$. This step will describe how the users change titles. We will report both the direction of change for statistically significant features and the Cohen's d effect size of that change. Since our features are roughly normally distributed, unlike the scores and number of comments, the Cohen's d effect size should be accurate.

\begin{table*}[htp!]
\centering
\caption{\label{tbl:ltr1}Evaluation of Lasso Regression models \textbf{Ranked by Score}. The test sets that are being ranked on range from 3000 to 8000 data points.}
\begin{tabular}{|c|c|c|c|c|c|c|c|c|c|c|c|} \hline
\multirow{2}{*}{Dataset} & \multirow{2}{*}{Model} & \multicolumn{5}{|c|}{Precision @ $k$} & \multicolumn{5}{|c|}{KT-distance @ $k$} \\ \cline{3-12}
& &$k=3$ & $k=10$  & $k=50$ & $k=100$ & $k=500$   &  $k=3$ & $k=10$ & $k=50$ & $k=100$ & $k=500$ \\ \hline
\multirow{1}{*}{r/worldnews 2012} & Senti+Subj &  1.0 & 0.9 & 0.78 & 0.78 & 0.76 & 3 & 26& 572 & 2300 & 62262\\
\hline
\multirow{1}{*}{r/worldnews 2013} & Senti+Subj & 0.677 & 0.667 & 00.80 & 0.78 & 0.78 & 3 & 27 & 703 & 2362 & 64177\\
\hline
\multirow{1}{*}{r/worldnews 2012} & Content &  1.0 & 1.0 & 0.84 & 0.77 & 0.84 & 2 & 20& 526 & 2063 & 59334\\
\hline
\multirow{1}{*}{r/worldnews 2013} & Content & 1.0 & 0.90 & 0.78 & 0.82 & 0.78 & 1 & 26 & 546 & 2776 & 62657\\
\hline
\multirow{1}{*}{r/worldnews 2012} & All &  1.0 & 1.0 & 0.88 & 0.87 & 0.86 & 3 & 30& 560 & 2553 & 60182\\
\hline
\multirow{1}{*}{r/worldnews 2013} & All & 1.0 & 0.70 & 0.80 & 0.78 & 0.71 & 4 & 18 & 559 & 2516 & 55869\\
\hline
\end{tabular}
\end{table*}

\begin{table*}[htp!]
\centering
\caption{\label{tbl:ltr_c}Evaluation of Lasso Regression models  \textbf{Ranked by \# of comments}, The test sets that are being ranked on range from 3000 to 8000 data points.}
\begin{tabular}{|c|c|c|c|c|c|c|c|c|c|c|c|} \hline
\multirow{2}{*}{Dataset} & \multirow{2}{*}{Model} & \multicolumn{5}{|c|}{Precision @ $k$} & \multicolumn{5}{|c|}{KT-distance @ $k$} \\ \cline{3-12}
& &$k=3$ & $k=10$ & $k=50$ & $k=100$ & $k=500$   &  $k=3$ & $k=10$ & $k=50$ & $k=100$ & $k=500$ \\ \hline
\multirow{1}{*}{r/worldnews 2012} & Senti+Subj &  1.0 & 1.0 & 0.88 & 0.87 & 0.86 & 3 & 30& 560 & 2553 & 60182\\
\hline
\multirow{1}{*}{r/worldnews 2013} & Senti+Subj & 1.0 & 0.80 & 0.80 & 0.78 & 0.734 & 6 & 16 & 643& 2689 & 58089\\
\hline
\multirow{1}{*}{r/worldnews 2012} & Content &  0.334 & 0.60 & 0.62 & 0.31 & 0.062 & 0 & 9& 656 & 656 & 656\\
\hline
\multirow{1}{*}{r/worldnews 2013} & Content & 0.66 & 0.20 & 0.66 & 0.60 & 0.25 & 1 & 1 & 482 & 1406 & 8476\\
\hline
\multirow{1}{*}{r/worldnews 2012} & All &  1.0 & 0.90 & 0.76 & 0.79 & 0.78 & 4 & 29& 596 & 2519 & 45293\\
\hline
\multirow{1}{*}{r/worldnews 2013} & All & 0.0 & 0.0 & 0.78 & 0.79 & 0.75 &  0& 0 & 68 & 1745 & 57661\\
\hline
\end{tabular}
\end{table*}

\section{Results and Discussion}

\subsection{Ranking News Popularity}\label{ranking results}
\textit{Please see attached erratum for updated results in this section}\\

First, we will present our findings on general news popularity on reddit. In Table~\ref{tbl:ltr1} and Table~\ref{tbl:ltr_c}, we report our results for each data set and several models. Table~\ref{tbl:ltr1} is reporting the results for ranking by score and Table~\ref{tbl:ltr_c} is reporting the results for ranking by the number of comments. 

When ranking by score, we find that the use of all of features (sentiment, subjectivity, and content) works best, achieving very high precision for all values of $k$. However, when separating the models, we find that content features predict slightly better than sentiment and subjectivity features.

When ranking by the number of comments, we find that the use of only sentiment and subjectivity features is best, achieving a much higher precision in values of small $k$ than content features. In fact, when adding content features to the model, we find a harsh decrease in performance, likely due to over-fitting or skewing the learning to less useful content features rather than more useful sentiment features. 

This difference in ranking performance suggests that the motivation of of the crowd in voting and commenting may be different. While these metrics of popularity are highly correlated, we do find distinct differences in the types of features that perform well in prediction. Predicting the score is based on both content and sentiment, while predicting the number of comments is based on sentiment only. This finding needs to be explored more fully, but is worth mentioning here. More importantly, we find that despite the salience of timing on reddit and in general news popularity, we are able to predict the crowds voting and engagement well without temporal features. This result both strengthens previous news popularity findings and illustrates the robustness of this feature set. 

To better understand these models, we will discuss our post-hoc analysis findings for three of our models: sentiment and subjectivity features for each ranking and all features combined for score ranking. Due to space constraints, we will only discuss these results, rather than display them in a table.

We find that many of the results in previous news popularity literature still hold with crowds on reddit. Specifically, we find that higher score news is more negative, more emotional overall, more certain, and focuses on the present. These findings align with~\cite{reis2015breaking}, ~\cite{harcup2001news}, and ~\cite{lewandowsky2012misinformation}. Further, we  find that higher score news is less subjective, displays more affiliation and  more clout. These findings make intuitive sense with what we know about traditional news organizations. Similarly, when ranking by comments, we find that news with a higher number of comments is more negative, but less emotional overall, more insightful, more certain, less subjective, and focuses on the present. 

When adding in content features, we find that higher score news is longer, uses more quotes, more personal pronoun references, and uses more simple, more readable words. These findings also align with previous literature, in particular with results in ~\cite{keneshloo2016predicting}. Interestingly, and somewhat counter-intuitively, we also see that articles With swear words in the original title from the news source tend to get higher scores. While this feature does not get very strong statistical significance on the two-class divide, it is used to predict in the models over both data sets. This may be due to the specific reddit community or a few high score outliers that contain swear words. Further investigation of this we leave to future work.

\begin{table}[t!]
\caption{Features that significantly differ between post titles and articles titles when title pair has less than $0.1$ cosine similarity. If a feature is significant in both data sets, we report the average effect size between the two, otherwise we report the effect size in the single data set. All results have Wilcox Rank-Sum P-values of at least less than 0.05.}
\label{BodyResults}
\hspace*{-0.2in}\begin{tabular}{  |c || c |c| c| }
\hline
 \textbf{Feature} & \textbf{r/worldnews 2012} & \textbf{r/worldnews 2013} & \textbf{Avg. Effect}\\
 \hline
 posemo & changed $>$ original & changed $>$ original & 0.0507\\
 negemo & changed $<$ original & changed $<$ original & 0.1497\\
 ttr & changed $>$ original &  & 0.1448\\
 affect & changed $<$ original & & 0.0922\\
 clout & changed $>$ original & & 0.0123\\
 swear & changed $>$ original & changed $>$ original & 0.0154\\
 informal & changed $>$ original & & 0.0705\\
 per\_stop & changed $>$ original & & 0.4634\\
 WC & changed $>$ original & changed $>$ original & 0.4457\\
 FKE & changed $>$ original & changed $>$ original & 0.3416\\
 word\_len & changed $>$ original & changed $>$ original & 0.1595\\
 tone & & changed $<$ original & 0.1001\\
 vad\_pos & & changed $>$ original & 0.0678\\
 netspeak & & changed $>$ original & 0.0616\\
 pos\_str & & changed $>$ original & 0.2147\\
 sixltr & & changed $>$ original & 0.0116\\
\hline
\end{tabular}
\end{table}
\begin{figure}[h]
\begin{center}
  \hspace*{-0.3in}\begin{tabular}{cc} \\
    {\small r/worldnews 2012} & {\small r/worldnews 2013} \\
    \includegraphics[width=130pt,keepaspectratio=true]{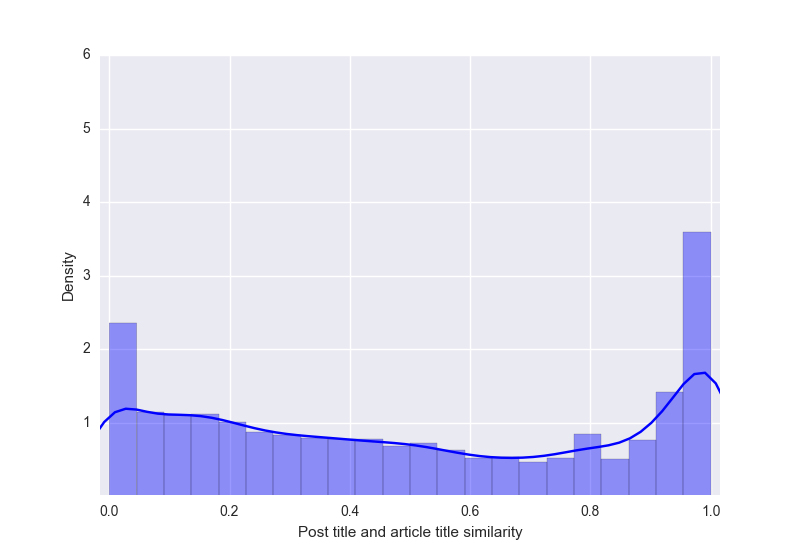}
    & \includegraphics[width=130pt,keepaspectratio=true]{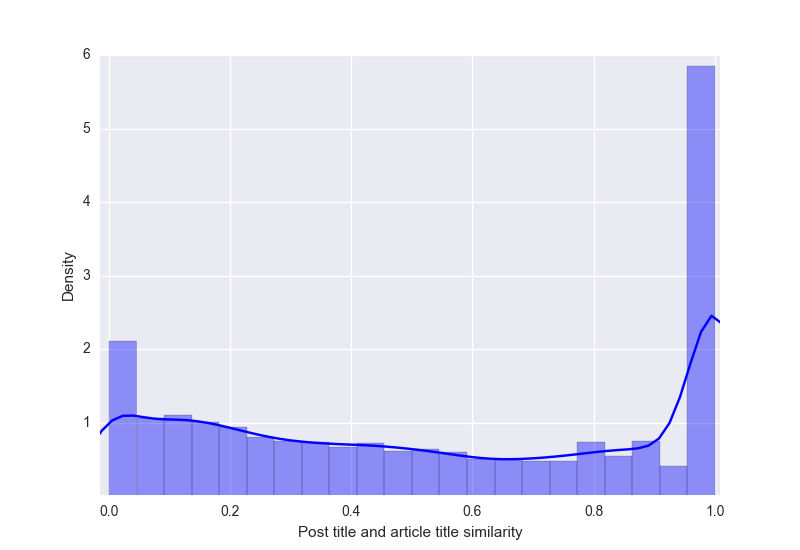} \\
    \multicolumn{2}{c}{(a) Title Similarity Distributions} \\
    \includegraphics[width=120pt,keepaspectratio=true]{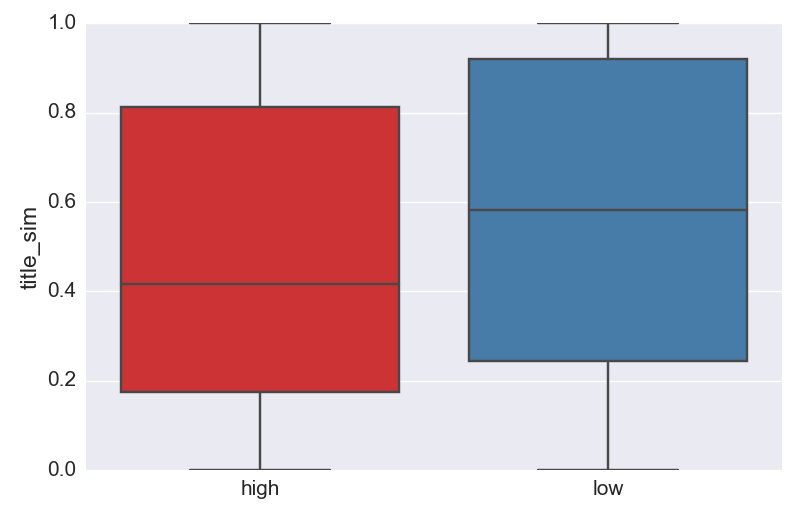}
    & \includegraphics[width=120pt,keepaspectratio=true]{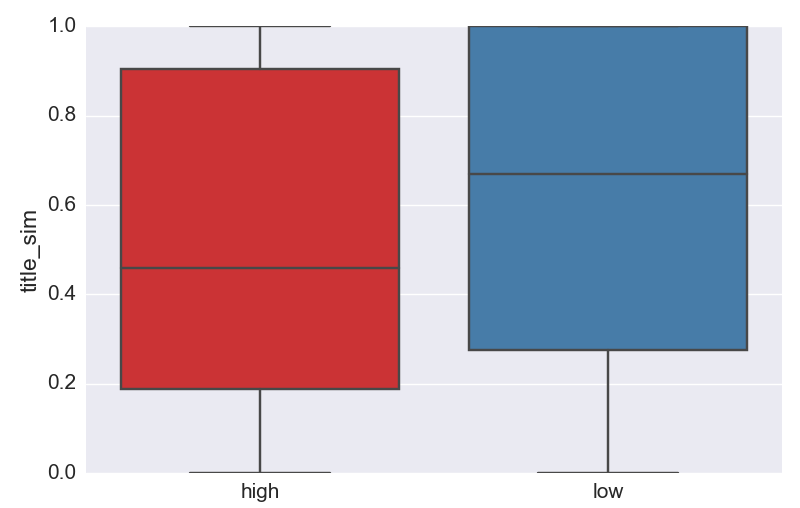}    
    \\
    \multicolumn{2}{c}{(b) High score v. Low score} \\
    \includegraphics[width=120pt,keepaspectratio=true]{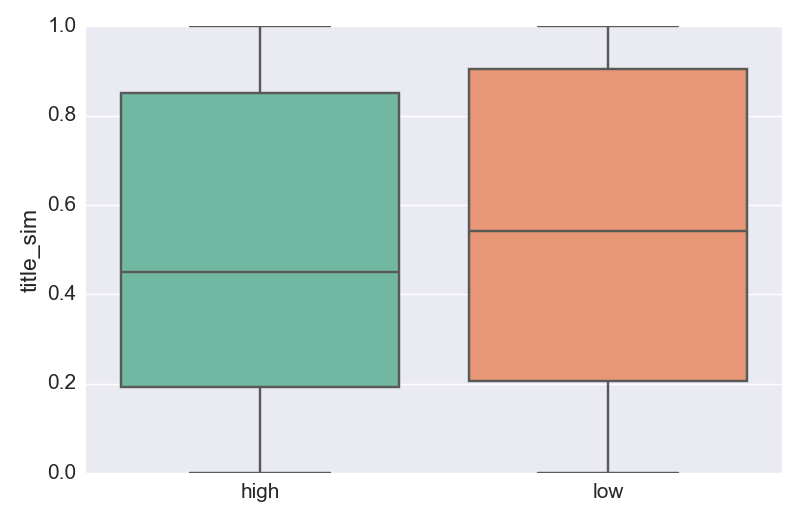}
    & \includegraphics[width=120pt,keepaspectratio=true]{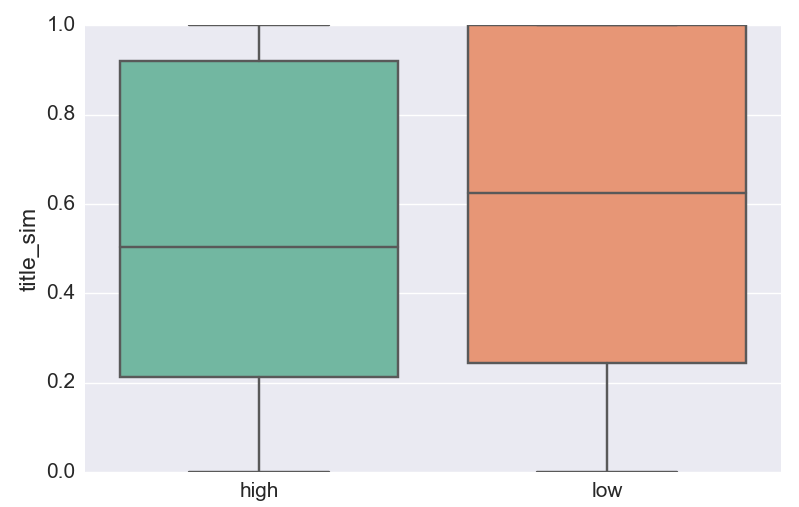}    
    \\
    \multicolumn{2}{c}{(c) High \# comments v. Low \# comments} \\
\end{tabular}
  \caption{(a) Distribution of cosine similarities between post title and article title. (b) Box plots of title similarity between top 90th percentile of scores and bottom 50th percentile of scores. (c) Box plots of title similarity between top 90th percentile of \# comments and bottom 50th percentile of \# comments. See Table~\ref{titlestats} for significance statistics.\label{titles}}
\end{center}
\end{figure}

\subsection{Title Changing}
Next, we will present some findings on title changing behavior. First, we find that most titles are changed with 85\% of titles being changed in 2012 and 71\% of titles being changed in 2013. In Figure~\ref{titles}a, we can see that the distribution of title similarities shows many titles being changed significantly. Saliently, we also find that changed titles tend to get a higher score than those that are not changed. In Table~\ref{titlestats}, we show that both the Wilcox Rank-Sum p-value and the Cohen's d effect size demonstrate a significant shift in the title similarity distributions of the top 90th percentile of scores and the bottom 50th percentile of scores. The corresponding box plots can be found in Figure~\ref{titles}. We find the same trend with the number of comments under a news post, but with less significance. It is critical to note that these significance values may be slightly inflated due to the skew in our score and number of comment distributions. To show this, we also compute the probability that random post in the high score category has a lower title similarity than a random post in the low score category. This method is meant to estimate effect size for nonparametric test~\cite{grissom2012effect}, but can be influenced by imbalanced sets as we have in our two-class divide. As expected, we get less significant effect sizes using this method, but still better than random chance. We can conclude that there is a significant difference in the distributions, but it may be less significant than our Cohen's d effect sizes show. As a whole, this is a surprising finding. It suggests that the crowd on reddit has a noticeable impact on news popularity and new engagement, which in turn, may change how people interpret the news and form public opinion.

To dive deeper into the behavior of title changing, we perform hypothesis testing on pairs of titles with cosine similarity less than $0.1$. We find that users change titles to be more informal, contain more swear words, more clout words, more difficult to read, and longer overall. Saliently, we find that original titles are significantly more negative than changed titles. These findings suggest that users are both adding their own analysis in the post title and providing a positive spin. This added positive analysis is in turn gaining slightly more popularity and engagement than the story may have with its original title. These results should be further inspected, but provide some insight into how crowds can change not just the popularity of news, but other users' perceptions when reading the news. These results and their corresponding effect sizes are shown in Table~\ref{BodyResults}.

\begin{table}[th]
\caption{Significance and effect size tests for the title similarities on a high vs low popularity split. Grissom-Kim probability is the probability that a random highly popular article has a lower title similarity than a random lowly popular article. See Figure~\ref{titles} for boxplots and distributions.}
\label{titlestats}
\hspace*{0in}\begin{tabular}{  |c |c || c |c|  }
\hline
 \textbf{data} & \textbf{Sig Metric} & \textbf{Score} & \textbf{\# Cmts}\\
 \hline
 \multirow{3}{*}{2012} & Wilcox P-Value & 7.736e-05 & 8.403e-05\\
 & Cohen's d Effect Size & 0.1557 & 0.1439\\
& Grissom-Kim Probability & 0.55 & 0.53\\
\hline 

\multirow{3}{*}{2013} & Wilcox P-Value & 8.194e-11 & 1.107e-13 \\
 &Cohen's d Effect Size &0.1368 & 0.1256 \\
 &Grissom-Kim Probability &0.54 &0.51 \\
 \hline
\end{tabular}
\end{table}

\section{Conclusions and Future Work}
In this paper, we present a preliminary data exploration of popular and highly engaged news on reddit. We found that many previous findings about news popularity still hold true on reddit, including stories with negative titles are more popular and stories that are more emotional overall are more popular. We also found that despite the confounding impact of time on reddit and news popularity in general, we are able to predict popularity reasonably well without temporal features. We find that content features can do reasonably well in predicting the votes of the community, but do not do well at predicting the length of discussion under the news article. On the other hand, sentiment features seem to be an excellent predictor of both the score and the length of discussion. In addition, we find that the crowd does have a noticable impact on news popularity and discussion through changing the article headline. When members of the crowds do change the headline, they add their own analysis and make the headline more positive than the news producers headlines. 

In the future, we want to extend this work in two ways. First, we want to explore more in depth popularity analysis and crowd effects across multiple platforms. For example, is a news article that is highly popular on reddit also popular on Facebook or Twitter? If this popularity differs, in what ways does it differ? Second, we would like to further investigate the impact changed news headlines. Specifically, we would like to expand this data set to include more reddit news communities and more social platforms over a longer period of time.

\section{Acknowledgments}
Research was sponsored by the Army Research Laboratory and was
accomplished under Cooperative Agreement Number W911NF-09-2-0053 (the
ARL Network Science CTA). The views and conclusions contained in this
document are those of the authors and should not be interpreted as
representing the official policies, either expressed or implied, of
the Army Research Laboratory or the U.S. Government. The
U.S. Government is authorized to reproduce and distribute reprints for
Government purposes notwithstanding any copyright notation here on. I would also like to thank Sujoy Sikdar for his learning to rank code

\bibliographystyle{aaai}
\bibliography{references}

\begin{thebibliography}{}

\bibitem[\protect\citeauthoryear{Bakshy, Messing, and
  Adamic}{2015}]{bakshy2015exposure}
Bakshy, E.; Messing, S.; and Adamic, L.~A.
\newblock 2015.
\newblock Exposure to ideologically diverse news and opinion on facebook.
\newblock {\em Science} 348(6239):1130--1132.

\bibitem[\protect\citeauthoryear{Davies}{2008 }]{coca}
Davies, M.
\newblock 2008-.
\newblock The corpus of contemporary american english: 520 million words,
  1990-present.
\newblock Available online at "\url{http://corpus.byu.edu/coca/}".
\newblock Accessed: 7/21/2015.

\bibitem[\protect\citeauthoryear{Ecker \bgroup et al\mbox.\egroup
  }{2014}]{ecker2014effects}
Ecker, U.~K.; Lewandowsky, S.; Chang, E.~P.; and Pillai, R.
\newblock 2014.
\newblock The effects of subtle misinformation in news headlines.
\newblock {\em Journal of experimental psychology: applied} 20(4):323.

\bibitem[\protect\citeauthoryear{Gilbert}{2013}]{gilbert2013widespread}
Gilbert, E.
\newblock 2013.
\newblock Widespread underprovision on reddit.
\newblock In {\em Proceedings of the 2013 conference on Computer supported
  cooperative work},  803--808.
\newblock ACM.

\bibitem[\protect\citeauthoryear{Grissom and Kim}{2012}]{grissom2012effect}
Grissom, R.~J., and Kim, J.~J.
\newblock 2012.
\newblock {\em Effect sizes for research: Univariate and multivariate
  applications}.
\newblock Routledge.

\bibitem[\protect\citeauthoryear{Harcup and O'neill}{2001}]{harcup2001news}
Harcup, T., and O'neill, D.
\newblock 2001.
\newblock What is news? galtung and ruge revisited.
\newblock {\em Journalism studies} 2(2):261--280.

\bibitem[\protect\citeauthoryear{Hessel, Lee, and Mimno}{2017}]{hessel2017cats}
Hessel, J.; Lee, L.; and Mimno, D.
\newblock 2017.
\newblock Cats and captions vs. creators and the clock: Comparing multimodal
  content to context in predicting relative popularity.
\newblock {\em arXiv preprint arXiv:1703.01725}.

\bibitem[\protect\citeauthoryear{Hessel, Tan, and
  Lee}{2016}]{hessel2016science}
Hessel, J.; Tan, C.; and Lee, L.
\newblock 2016.
\newblock Science, askscience, and badscience: On the coexistence of highly
  related communities.
\newblock In {\em Tenth International AAAI Conference on Web and Social Media}.

\bibitem[\protect\citeauthoryear{Horne, Adal{\i}, and
  Chan}{2016}]{horne2016impact}
Horne, B.~D.; Adal{\i}, S.; and Chan, K.
\newblock 2016.
\newblock Impact of message sorting on access to novel information in networks.
\newblock In {\em Advances in Social Networks Analysis and Mining (ASONAM),
  2016 IEEE/ACM International Conference on},  647--653.
\newblock IEEE.

\bibitem[\protect\citeauthoryear{Horne, Adali, and
  Sikdar}{2017}]{horne2017identifying}
Horne, B.~D.; Adali, S.; and Sikdar, S.
\newblock 2017.
\newblock Identifying the social signals that drive online discussions: A case
  study of reddit communities.
\newblock {\em arXiv preprint arXiv:1705.02673}.

\bibitem[\protect\citeauthoryear{Horne \bgroup et al\mbox.\egroup
  }{2016}]{horne2016expertise}
Horne, B.~D.; Nevo, D.; Freitas, J.; Ji, H.; and Adali, S.
\newblock 2016.
\newblock Expertise in social networks: How do experts differ from other users?
\newblock In {\em Tenth International AAAI Conference on Web and Social Media}.

\bibitem[\protect\citeauthoryear{Hutto and Gilbert}{2014}]{hutto2014vader}
Hutto, C.~J., and Gilbert, E.
\newblock 2014.
\newblock Vader: A parsimonious rule-based model for sentiment analysis of
  social media text.
\newblock In {\em Eighth International AAAI Conference on Weblogs and Social
  Media}.

\bibitem[\protect\citeauthoryear{Jaech \bgroup et al\mbox.\egroup
  }{2015}]{jaech2015talking}
Jaech, A.; Zayats, V.; Fang, H.; Ostendorf, M.; and Hajishirzi, H.
\newblock 2015.
\newblock Talking to the crowd: What do people react to in online discussions?
\newblock {\em arXiv preprint arXiv:1507.02205}.

\bibitem[\protect\citeauthoryear{Keneshloo \bgroup et al\mbox.\egroup
  }{2016}]{keneshloo2016predicting}
Keneshloo, Y.; Wang, S.; Han, E.-H.; and Ramakrishnan, N.
\newblock 2016.
\newblock Predicting the popularity of news articles.
\newblock In {\em Proceedings of the 2016 SIAM International Conference on Data
  Mining},  441--449.
\newblock SIAM.

\bibitem[\protect\citeauthoryear{Lakkaraju, McAuley, and
  Leskovec}{2013}]{lakkaraju2013s}
Lakkaraju, H.; McAuley, J.~J.; and Leskovec, J.
\newblock 2013.
\newblock What's in a name? understanding the interplay between titles,
  content, and communities in social media.
\newblock {\em ICWSM} 1(2):3.

\bibitem[\protect\citeauthoryear{Lewandowsky \bgroup et al\mbox.\egroup
  }{2012}]{lewandowsky2012misinformation}
Lewandowsky, S.; Ecker, U.~K.; Seifert, C.~M.; Schwarz, N.; and Cook, J.
\newblock 2012.
\newblock Misinformation and its correction continued influence and successful
  debiasing.
\newblock {\em Psychological Science in the Public Interest} 13(3):106--131.

\bibitem[\protect\citeauthoryear{Pang and Lee}{2004}]{pang2004sentimental}
Pang, B., and Lee, L.
\newblock 2004.
\newblock A sentimental education: Sentiment analysis using subjectivity
  summarization based on minimum cuts.
\newblock In {\em Proceedings of the 42nd annual meeting on Association for
  Computational Linguistics},  271.
\newblock Association for Computational Linguistics.

\bibitem[\protect\citeauthoryear{Pedregosa \bgroup et al\mbox.\egroup
  }{2011}]{pedregosa2011scikit}
Pedregosa, F.; Varoquaux, G.; Gramfort, A.; Michel, V.; Thirion, B.; Grisel,
  O.; Blondel, M.; Prettenhofer, P.; Weiss, R.; Dubourg, V.; et~al.
\newblock 2011.
\newblock Scikit-learn: Machine learning in python.
\newblock {\em Journal of Machine Learning Research} 12(Oct):2825--2830.

\bibitem[\protect\citeauthoryear{Petty and Cacioppo}{1986}]{petty1986}
Petty, R.~E., and Cacioppo, J.~T.
\newblock 1986.
\newblock The elaboration likelihood model of persuasion.
\newblock In {\em In Communication and Persuasion}. New York: Springer.
\newblock  1--24.

\bibitem[\protect\citeauthoryear{Reis \bgroup et al\mbox.\egroup
  }{2015}]{reis2015breaking}
Reis, J.; Benevenuto, F.; de~Melo, P.~O.; Prates, R.; Kwak, H.; and An, J.
\newblock 2015.
\newblock Breaking the news: First impressions matter on online news.
\newblock {\em arXiv preprint arXiv:1503.07921}.

\bibitem[\protect\citeauthoryear{Surber and Schroeder}{2007}]{surber2007effect}
Surber, J.~R., and Schroeder, M.
\newblock 2007.
\newblock Effect of prior domain knowledge and headings on processing of
  informative text.
\newblock {\em Contemporary educational psychology} 32(3):485--498.

\bibitem[\protect\citeauthoryear{Tan and Lee}{2015}]{tan2015all}
Tan, C., and Lee, L.
\newblock 2015.
\newblock All who wander: On the prevalence and characteristics of
  multi-community engagement.
\newblock In {\em Proceedings of the 24th International Conference on World
  Wide Web},  1056--1066.
\newblock ACM.

\bibitem[\protect\citeauthoryear{Tausczik and
  Pennebaker}{2010}]{tausczik2010psychological}
Tausczik, Y.~R., and Pennebaker, J.~W.
\newblock 2010.
\newblock The psychological meaning of words: Liwc and computerized text
  analysis methods.
\newblock {\em Journal of language and social psychology} 29(1):24--54.

\bibitem[\protect\citeauthoryear{Thelwall \bgroup et al\mbox.\egroup
  }{2010}]{thelwall2010sentiment}
Thelwall, M.; Buckley, K.; Paltoglou, G.; Cai, D.; and Kappas, A.
\newblock 2010.
\newblock Sentiment strength detection in short informal text.
\newblock {\em Journal of the American Society for Information Science and
  Technology} 61(12):2544--2558.

\bibitem[\protect\citeauthoryear{Tran and
  Ostendorf}{2016}]{tran2016characterizing}
Tran, T., and Ostendorf, M.
\newblock 2016.
\newblock Characterizing the language of online communities and its relation to
  community reception.
\newblock {\em arXiv preprint arXiv:1609.04779}.

\bibitem[\protect\citeauthoryear{Tsagkias, Weerkamp, and
  De~Rijke}{2009}]{tsagkias2009predicting}
Tsagkias, M.; Weerkamp, W.; and De~Rijke, M.
\newblock 2009.
\newblock Predicting the volume of comments on online news stories.
\newblock In {\em Proceedings of the 18th ACM conference on Information and
  knowledge management},  1765--1768.
\newblock ACM.

\bibitem[\protect\citeauthoryear{Zaman \bgroup et al\mbox.\egroup
  }{2014}]{zaman2014bayesian}
Zaman, T.; Fox, E.~B.; Bradlow, E.~T.; et~al.
\newblock 2014.
\newblock A bayesian approach for predicting the popularity of tweets.
\newblock {\em The Annals of Applied Statistics} 8(3):1583--1611.

\end{thebibliography}


\begin{table*}[h]
\centering
\caption{\label{corr1}Sample of corrected ranking evaluations \textbf{Ranked by Score}.}
\begin{tabular}{|c|c|c|c|c|c|c|c|c|c|c|c|} \hline
\multirow{2}{*}{Dataset} & \multirow{2}{*}{Model} & \multicolumn{5}{|c|}{Precision @ $k$} & \multicolumn{5}{|c|}{KT-distance @ $k$} \\ \cline{3-12}
& &$k=3$ & $k=10$  & $k=50$ & $k=100$ & $k=500$   &  $k=3$ & $k=10$ & $k=50$ & $k=100$ & $k=500$ \\ \hline

\multirow{1}{*}{r/worldnews 2013} & Senti+Subj & 0.0 & 0.0 & 0.0 & 0.0 & 0.0 & 2 & 13 & 341 & 1748 & 42907\\
\hline
\multirow{1}{*}{r/worldnews 2013} & Content & 0.0 & 0.0 & 0.02 & 0.04 & 0.14 & 5 & 23 & 602 & 2562 & 61927\\
\hline
\multirow{1}{*}{r/worldnews 2013} & All & 0.0 & 0.0 & 0.03 & 0.04 & 0.13 & 3 & 23 & 597 & 2446 & 63668\\
\hline
\end{tabular}
\end{table*}

\begin{table*}[h]
\centering
\caption{\label{corr2}Sample of corrected ranking evaluations \textbf{Ranked by \# of comments}.}
\begin{tabular}{|c|c|c|c|c|c|c|c|c|c|c|c|} \hline
\multirow{2}{*}{Dataset} & \multirow{2}{*}{Model} & \multicolumn{5}{|c|}{Precision @ $k$} & \multicolumn{5}{|c|}{KT-distance @ $k$} \\ \cline{3-12}
& &$k=3$ & $k=10$  & $k=50$ & $k=100$ & $k=500$   &  $k=3$ & $k=10$ & $k=50$ & $k=100$ & $k=500$ \\ \hline

\multirow{1}{*}{r/worldnews 2013} & Senti+Subj & 0.0 & 0.0 & 0.0 & 0.0 & 0.0 & 0 & 0 & 0 & 6 & 53\\
\hline
\multirow{1}{*}{r/worldnews 2013} & Content & 0.0 & 0.0 & 0.06 & 0.04 & 0.12 & 5 & 26 & 686 & 2612 & 59950\\
\hline
\multirow{1}{*}{r/worldnews 2013} & All & 0.0 & 0.0 & 0.00 & 0.02 & 0.09 & 1 & 27 & 566 & 2406 & 60885\\
\hline
\end{tabular}
\end{table*}
\section{Erratum}
\textit{Date of Erratum Posting:11/3/2017}\\

After subsequent research, it was found that the learning to rank results in Tables \ref{tbl:ltr1} and \ref{tbl:ltr_c} are inflated due to a bug in the original ranking code.  Thus, in Tables ~\ref{corr1} and ~\ref{corr2} we display a sample of the ranking results, with the bug fixed, for both ranking by score and ranking by the number of comments. These new results invalidate our previous claim that we are able to predict reddit post ranking reasonably well without temporal features. 

To investigate further, we attempt to classify the top 5\% of news posts based on the score and the bottom 5\% of news posts based on the score using a random forest classifier. We can see that there is indeed some signal, as we achieve 67\% accuracy over a 53\% baseline. However, there does not seem to be enough signal to gain reasonable predictions when including much of the middle ranked posts. Looking at the current literature, this finding makes sense. In ~\cite{hessel2017cats}, it is discussed that both time and the ''rich-get-richer" dynamics can create complex and non-linear relationships between the quality of a post and the popularity. This finding should also hold true with news posts, perhaps even more so, as the attention in the news cycle changes in a complex way due to external events that our features have no way of capturing. Thus, we would presumably need a much more complex set of features that capture both time and external attention to accurately predict the ranking of news on reddit.

On the other hand, as shown in ~\cite{horne2017identifying} and ~\cite{jaech2015talking}, a similar ranking technique can perform well in ranking the popularity of comments under a post, including news posts. This finding is likely due to the limited context and limited time-of-life of a single post, allowing the overall quality based on language and sentiment features to be valid. In ~\cite{horne2017identifying}, we still see that time has an impact on popularity, but not consistently.

While our previous ranking results do not hold, all other results in this paper do hold, including the feature introspection on the two-class divide, and our title change study.

\end{document}